\documentclass[a4paper,12pt]{article}

\newtheorem{Prop}{Proposition}

\newtheorem{Teo}[Prop]{Theorem}
\newtheorem{Lem}[Prop]{Lemma}

\newcommand{\D}{\mathrm{d}}
\newcommand{\BE}{\begin{equation}}
\newcommand{\EE}{\end{equation}}

\newcommand{\T}[1]{\mathbf{#1}}

\begin{document}

\title{Complex variables for separation of Hamilton-Jacobi equation on three-dimensional Minkowski space}

\author{
Luca Degiovanni\footnote{luca.degiovanni@gmail.org}, 
Giovanni Rastelli\footnote{giorast@tin.it}\\
Dipartimento di Matematica, Universit\`a di Torino,\\ Via Carlo Alberto 10,
10123 Torino, Italy
}
\date{}

\maketitle

\begin{abstract}
The real coordinates separating geodesic Hamilton-Jacobi equation on three-dimensional Minkowski space in several cases cannot be defined in the whole space. We show through an example how to naturally extend them to complex variables defined everywhere (excluding the singular surfaces of each coordinate system only) and still separating the same equation. 
\end{abstract}


\section{\large Introduction}

The separability of the geodesic Hamilton-Jacobi (HJ) equation in some orthogonal coordinate system in $n$-dimensional Riemannian or pseudo-Riemannian manifolds is characterized by a Killing-St\"ackel (KS) algebra of $n$ (including the metric tensor) independent, commuting, simultaneously diagonalized Killing two-tensors  with common closed eigenforms that define the separable hypersurfaces (see \cite{Ben} and references therein).

On the three-dimensional Euclidean space eleven such separable coordinate systems exist, classified first in \cite{E}. On the three-dimensional Minkowski space many more exist, up to 82 according to \cite{Hi}. Differently from the euclidean case, several of these systems are undefined in open subsets of the whole space, due to the fact that eigenforms of the KS-algebra can be there complex valued (the same happens in two-dimensional Minkowski space). In \cite{complex} we define complex variables for this cases  and develop a geometric theory of separation for HJ equation in  complex variables on real pseudo-Riemannian manifolds, with exhaustive examples for the two-dimensional Minkowski space. In the present paper we introduce complex separable variables on three-dimensional Minkowski space through a meaningful example and by making use of results exposed in \cite{CRCG} and \cite{arkiv} where separable coordinates are directly related to the eigenvalues of a basis of their KS algebra. These variables fit naturally with the standard real separable coordinate systems. A complete setting of complex separable variables on three-dimensional Minkowski space is still to be done.  In Sections 2 and 3 we recall shortly the results of \cite{complex}, Section 4 is devoted to the example.

\section{\large Definitions and first results}
A tensor $\T{K}$ on a real $n$-dimensional pseudo-Riemannian manifold $Q$ (with real coordinates ($q^i$) and metric tensor $\T{g}$) is called a \emph{Killing tensor} if its covariant components satisfy the equation $\nabla_{(k} K_{i,\cdots,j)}=0$. A classical characteristic Killing tensor is a Killing two-tensor $\T{K}$ that (in a open subset of $Q$) admits $n$ pointwise distinct real functions $(\lambda^i)$ and $n$ functionally independent real functions $(z^i)$ that satisfy (with no summation over $i$) the equations:
\BE\label{autovet}
\T{K}\D z^i =\lambda^i \T{g}\D z^i\,.
\EE
Locally, to any characteristic Killing tensor can be associated a KS-algebra, and vice-versa.

In the paper~\cite{complex} it was shown how to drop the requirement that the functions $(\lambda^i)$ and $(z^i)$ are real and how the St\"ackel-Eisenhart theory and the Levi-Civita criterion can be extended to this case, without to complexify the manifold $Q$.

Because $\T{K}$ and $\T{g}$ are both symmetric with real coefficients,  if $z^i$ satisfies the relation~(\ref{autovet}) also its complex conjugate $\bar{z}^i$ satisfies the same relation. Hence, the functions $(z^i)$ with a non-vanishing imaginary part arise in pairs and the $n$ complex-valued functions $(z^i)$ define, through their real and imaginary parts, $n$ real-valued coordinates.

It is well known that, if $(p_i)$ are the conjugate momenta of the coordinates $(q^i)$ on $T^\ast Q$, one can define the the momenta conjugate to the variables $z^i$ by using the Jacobian matrix $\left(\frac{\partial z^i}{\partial q^j}\right)$ and its inverse $\T{J}$ with components $J^j_i$:
\BE\label{trasf_inv}
P_j=J^i_j p_i\,.
\EE

\begin{Lem}
The relations \mbox{$\{P_i,z^j\}=\delta^j_i$}, \mbox{$\{z^i,z^j\}=0$} and \mbox{$\{P_i,P_j\}=0$} hold, by extending the Poisson bracket to complex function in the natural way. Therefore the variables \mbox{$(P_i$, $z^i)$} are canonical.
\end{Lem}
Moreover, to the variables $(z^i,P_i)$ are associated the, eventually complex-valued, vector fields: 
\begin{eqnarray}
Z_i=\frac{\partial}{\partial z^i}&=&J^j_i\frac{\partial}{\partial q^j} \label{zder}\\
\frac{\partial}{\partial P_i}&=&\frac{\partial z^i}{\partial q^j}\frac{\partial}{\partial p_j} \label{Pder}
\end{eqnarray}

\begin{Lem}\label{Lem_der}
The vector fields~(\ref{zder}) and (\ref{Pder}) satisfy the properties:
\begin{enumerate}
\item $Z_i(q^j)=J^j_i$ and $\overline{Z}_i(\overline{W})=\overline{Z_i(W)}$;
\item $Z_i(z^j)=\delta^j_i$ and if $\bar{z}^i\neq z^i$ then $Z_i(\bar{z}^i)=0$;
\item $\frac{\partial}{\partial \bar{z}^{i}}=\overline{\frac{\partial}{\partial z^i}}$ and $\frac{\partial}{\partial \bar{P}_{i}}=\overline{\frac{\partial}{\partial P_i}}$;
\item $\left[\frac{\partial}{\partial z^i},\frac{\partial}{\partial z^j}\right]=\left[\frac{\partial}{\partial P_i},\frac{\partial}{\partial P_j}\right]=\left[\frac{\partial}{\partial z^i},\frac{\partial}{\partial P_j}\right]=0$.
\end{enumerate}
\end{Lem}
\begin{Prop}
The Poisson bracket of two, possibly complex-valued, functions $f(z^i,P_i)$ and $g(z^i,P_i)$ is  given by
$$
\{f,g\}=\frac{\partial f}{\partial P_i}\frac{\partial g}{\partial z^i}-
\frac{\partial f}{\partial z^i}\frac{\partial g}{\partial P_i}\,.
$$
Hence, the Hamilton equations for an Hamiltonian $H(z^i,P_i)$ are
$$
\left\{\begin{array}{rcl}
\dot{z}^i&=&\frac{\partial H}{\partial P_i} \\[10pt]
\dot{P}_i&=&-\frac{\partial H}{\partial z^i}
\end{array}
\right.
$$
The Hamilton-Jacobi equation for the Hamiltonian $H(z^i,P_i)$ is
\begin{equation}\label{HJ}
H(z^i,\frac{\partial W}{\partial z^i})=h\,,
\end{equation}
where $W$ is, in general, a complex-valued function.
\end{Prop}

\section{\large Separation of variables for Hamilton--Jacobi equation}
A \emph{complete integral} of the equation~(\ref{HJ}) is a solution $W(z^i,c_j)$ of (\ref{HJ}), where $(c_j)$ are $n$ real-valued parameters, satisfying
$$
\det \left(\frac{\partial^2 W}{\partial z^i \partial c_j}\right)\neq0.
$$
A complete integral $W(z^i,c_j)$ is (additively) separated in the variables $(z^i)$ if
$$
W(z^i,c_j)=W_1(z^1,c_j)+\dots + W_n(z^n,c_j).
$$
In the paper~\cite{complex} is shown that an analogous of the Levi-Civita criterion~\cite{LC} holds also in the complex case:  
\begin{Teo}
The Hamilton--Jacobi equation for Hamiltonian $H$ admits a complete integral separated in the variables $(z^i)$, if and only if the Hamiltonian $H$ satisfies the Levi--Civita equation
$$
\partial^i H \partial^j H \partial_{ij} H+
\partial_i H \partial_j H \partial^{ij} H-
\partial^i H \partial_j H \partial_i^j H-
\partial_i H \partial^j H \partial^i_j H=0
$$
where $i\neq j$, no summation over the repeated indices is understood and
$$
\partial_i=\frac{\partial}{z^i}=\frac{\partial q^j}{\partial z^i}\frac{\partial}{\partial q^j}\,,
\quad
\partial^i=\frac{\partial}{P_i}=\frac{\partial z^i}{\partial q^j}\frac{\partial}{\partial p_j}\,.
$$
\end{Teo}
\begin{Prop}
If the Hamiltonian $H$ is real and satisfies the Levi--Civita equations then the complete integral can be chosen to be real.
\end{Prop}

Given a basis $(\T{K}_{k})$ of a KS-algebra, and denoted by $\lambda_{k}^i$ the eigenvalues of $\T{K}_{k}$ then, because the $(\T{K}_{k})$ are simultaneously diagonalized, the HJ equations associated to them can be set in the matrix form
\BE\label{Systemsep}
\left(\begin{array}{ccc}
\lambda_1^1g^1 & \cdots & \lambda_1^ng^n \\
\vdots && \vdots \\
\lambda_{n-1}^1g^1 & \cdots & \lambda_{n-1}^ng^n \\
g^1 & \cdots & g^n
\end{array}\right)
\left(\begin{array}{c}
(\partial_1W)^2\\ (\partial_2W)^2\\ \vdots\\ (\partial_nW)^2
\end{array}\right)
=
\left(\begin{array}{c}
c_{1}\\ c_{2}\\ \vdots\\ c_{n}
\end{array}\right).
\EE
The equations obtained by inverting (\ref{Systemsep}) admit complete separated solutions if and only if, denoting by $\T{S}$  the inverse of the matrix in the left side of (\ref{Systemsep}), the $i$-th row of $\T{S}$ depends on the variable $z^i$ only. Such a matrix is called \emph{St\"ackel matrix}. 

Classical results state that:
\begin{Teo}
The functions $K_{k}=\frac{1}{2}\sum_i\lambda_{k}^i g^i{P_i}^2$ e $H=\frac{1}{2}\sum_j g^j {P_j}^2$ are in involution if and only if the Killing-Eisenhart conditions hold:
\begin{equation}\label{Eisenhart}
g^j\frac{\partial\lambda_{k}^j}{\partial z^i}=(\lambda_{k}^i-\lambda_{k}^j)\frac{\partial g^j}{\partial z^i}\,.
\end{equation}
\end{Teo}
\begin{Teo}
The Eisenhart equations (\ref{Eisenhart}) hold if and only if the entries of the invertible matrix $\T{S}$ satisfy $\frac{\partial}{\partial z^i} S^k_j=0$ for $i\neq j$, i.e. it is a St\"ackel matrix.
\end{Teo}
In \cite{complex} we show that the previous result holds also in the complex setting.

\section{\large Example}
Orthogonal separable coordinates for HJ equation in the three-dimensional Minkowski space are described and classified in several papers by Kalnins and Miller \cite{Kal}, \cite{KM} and more details are given in \cite{Hi} where the separable systems are defined according to the scheme given in \cite{KM} and their singular sets are shown. The papers \cite{CRCG}, \cite{arkiv}  show how the real eigenvalues of any basis of a Killing-St\"ackel algebra can be combined together to build functions depending each on one only of the associated non-ignorable separable coordinated and that, in the case when all that functions are constants or undefined, the corresponding separable coordinate is ignorable. In the three-dimensional case, if $(\lambda_1^i)$ and $(\lambda_2^i)$ are the eigenvalues of a basis of a KS-algebra, the relevant functions depending on the single separable coordinate $z^h$ are 
\begin{equation}
f^{21}_h=\frac{\lambda_1^i-\lambda_1^j}{\lambda_2^i-\lambda_2^j} \qquad
f^{3k}_h=(-1)^{3+k}\frac{\lambda_1^i\lambda_2^j-\lambda_1^j\lambda_2^i}{\lambda_2^i-\lambda_2^j}
\label{funz}
\end{equation}
with $h,k=1,\dots ,3$ and $i=h+1$, $j=h+2$ mod 3.
In this paper we consider here separable systems without ignorable coordinates (asymmetric systems) only. It is easy to see from \cite{CRCG}, \cite{arkiv} that  the functions  (\ref{funz}) behave in the same way also in the complex case, being the proof of this property based on the Killing-Eisenhart equations only (see the previous section). Therefore, the functions (\ref{funz}) depend on only one of the separable variables, real or complex, and can be used to define exactly this variable. For example, let us consider in the Minkowski three-dimensional space the real systems $(\mu,\nu,\rho)$ (the (B.1.c) in \cite{Hi}, see also the (B.1.d) in the same paper, only slightly different) defined in pseudo-cartesian coordinates $(t,x,y)$ by

\begin{eqnarray}
t^2&=&\frac{\mu\nu\rho}a \nonumber\\
x^2&=&\frac{(\mu-a)(a-\nu)(\rho-a)}{a(a-1)}\label{def}\\
y^2&=&\frac{(\mu-1)(\nu-1)(\rho-1)}{a-1} \nonumber
\end{eqnarray}
where inequalities involving $(\mu,\nu,\sigma)$ and $0,1,a,$ make the distinction among systems defined in some open sets of the space separated by singular surfaces, ``horizons" in \cite{Hi}, and assure here the reality of the greek variables that remain undefined in some other open sets. 

The components of the metric tensor in the coordinates $(\mu,\nu,\sigma)$ are given by
$$
g^1=  - \displaystyle \frac {4\,(\mu  - 1)\,(\mu  - a
)\,\mu }{\nu \,\rho  + \mu ^{2} - \mu \,\nu  - \rho \,\mu } \quad 
g^2= \displaystyle \frac {4\,(\nu  - 1)\,( - a + \nu 
)\,\nu }{ - \rho \,\mu  - \nu ^{2} + \mu \,\nu  + \nu \,\rho } 
$$
$$
g^3=  - \displaystyle \frac {4\,(\rho  - 1)\,(\rho 
 - a)\,\rho }{\mu \,\nu  + \rho ^{2} - \rho \,\mu  - \nu \,\rho}. 
$$

To all these  systems corresponds the KS algebra generated by the metric $\T{K}_3=\T{g}$ and the Killing  tensors $\T{K}_1$ and $\T{K}_2$ with components, in $(t,x,y)$ coordinates: 
$$
K_1^{11}=x^2+y^2a+y^2-a, \quad K_1^{22}=t^2, \quad K_1^{33}=t^2a,
$$
$$
K_1^{12}=tx, \quad K_1^{13}=aty, \quad K_1^{23}=0,
$$
and
$$
K_2^{11}=x^2+y^2-a-1, \quad K_2^{22}=-y^2+t^2+1,\quad K_2^{33}=-x^2+t^2+a,
$$
$$
K_2^{12}=tx,\quad K_2^{13}=ty, \quad K_2^{23}=xy,
$$
whose eigenvalues $\lambda_1^i$ and $\lambda_2^i$, where  $(\mu, \nu,\rho)$ are defined, are given by

 $$
 \lambda_1^1=\nu \rho,  \quad \lambda_1^2=\mu \rho,  \quad \lambda_1^3=\mu \nu,
 $$
and
 $$
\lambda_2^1=\nu  + \rho, \quad \lambda_2^2=\mu  + \rho, \quad \lambda_2^3=\mu  + \nu ,
$$
respectively (we remark ``en passant" that the eigenvalues of above reveal a structure analogue to that of a Benenti system \cite{sb}, not yet studied in pseudo-Riemannian manifolds).  It is possible to see that the singular surfaces coincide with the points where the dimension of the KS algebra is $<3$, i.e. where 
$$
 \det \left( \begin{array}{ccc} \lambda_1^1&\lambda_1^2&\lambda_1^3 \\
 \lambda_2^1&\lambda_2^2&\lambda_2^3 \\ 
 1&1&1 \end{array} \right)=0 
$$
and that in the portions of the space where (\ref{def}) are undefined the KS algebra has a couple of complex conjugated eigenforms and eigenvalues.
The functions (\ref{funz}) provide the link between the eigenvalues and the separable coordinates, indeed, as instance,
\begin{eqnarray}
f^{21}_1&=&\frac{\lambda_1^2-\lambda_1^3}{\lambda_2^2-\lambda_2^3}=\mu \nonumber\\
f^{21}_2&=&\frac{\lambda_1^1-\lambda_1^3}{\lambda_2^1-\lambda_2^3}=\nu \label{A}\\
f^{21}_3&=&\frac{\lambda_1^1-\lambda_1^2}{\lambda_2^1-\lambda_2^2}=\rho \nonumber
\end{eqnarray}
We consider now the case when some of the common eigenforms of $\T{K}_1$ and $\T{K}_2$ are complex. If, for example, $\lambda_1^1=\bar{\lambda}_1^2$, $\lambda_2^1=\bar{\lambda}_2^2$ and $\lambda_1^3$, $\lambda_2^3$ are real we have 
$f^{21}_1=\bar{f}^{21}_2$ and $f^{21}_3$ real. These functions fit with the complex variables that we introduced in the previous sections and via (\ref{A}) extend naturally the $(\mu,\nu,\rho)$ variables in the whole Minkowski space deprived of the singular surfaces, having consequently,  $\mu=\bar{\nu}$ and $\rho$ real. 
It is remarkable that the $f^{21}_i$ and (\ref{def}) define always the same variables $(\mu,\nu,\rho)$ no matter if they are real or complex. In conclusion, the complex variables introduced above not only allow complete separation of variables for HJ equation where it is impossible to define real separable coordinates, but also represent the natural extension of the latter to the whole space (apart the singular surfaces). The mechanism of separation of variables and the integration of the separate complete integral of the HJ equations using complex variables are described in \cite{complex}.

\end{document}